%% file: main.tex
\documentclass[camera]{jpaper}
\usepackage{makecell}

\usepackage[nocompress]{cite}

\usepackage{soul}

\usepackage{amsmath}
\usepackage{graphicx}
\usepackage[linesnumbered,ruled]{algorithm2e}

\usepackage{algpseudocode}
\usepackage{titlesec}
\algrenewcommand\algorithmiccomment[2][\normalsize]{{#1\hfill\(\triangleright\) #2}}
\usepackage{geometry} \titlespacing*{\section}{0pt}{2pt plus 0.5pt minus 0.5pt}{0pt}
\titlespacing*{\subsection}{0pt}{2pt plus 0.5pt minus 0.5pt}{0pt}
\titlespacing*{\subsubsection}{0pt}{2pt plus 0.5pt minus 0.5pt}{0pt}

\makeatletter
\renewcommand{\@makefnmark}{\hbox{\textsuperscript{\scriptsize{\@thefnmark}}}}
\makeatother

\usepackage{array}
\newcolumntype{L}[1]{>{\raggedright\let\newline\\\arraybackslash\hspace{0pt}}m{#1}}
\newcolumntype{C}[1]{>{\centering\let\newline\\\arraybackslash\hspace{0pt}}m{#1}}
\newcolumntype{R}[1]{>{\raggedleft\let\newline\\\arraybackslash\hspace{0pt}}m{#1}}

\usepackage{bm}

\makeatletter
\let\MYcaption\@makecaption
\makeatother

\usepackage[font=footnotesize]{subcaption}

\makeatletter
\let\@makecaption\MYcaption
\makeatother

\usepackage{fixltx2e}

\usepackage{float}

\usepackage{dblfloatfix}
\usepackage[nolessnomore, italic]{mathastext}
\usepackage[T1]{fontenc}
\usepackage[usenames,dvipsnames,svgnames,table]{xcolor}
\usepackage{multirow}
\usepackage{hhline}
\usepackage[normalem]{ulem}
\usepackage{setspace}
\usepackage{indentfirst}
\usepackage{footmisc}

\usepackage{pifont}

\newif\ifcameraready
\camerareadyfalse

\definecolor{darkpink}{rgb}{0.88, 0.28, 0.54}

\newcommand{\affilETH}[0]{\textsuperscript{1}}
\newcommand{\affilBilkent}[0]{\textsuperscript{4}}
\newcommand{\affilBionano}[0]{\textsuperscript{3}}

\definecolor{amber}{rgb}{1.0, 0.49, 0.0}
\definecolor{darkgreen}{rgb}{0.0, 0.2, 0.13}
\definecolor{darkbyzantium}{rgb}{0.36, 0.22, 0.33}
\definecolor{darkseagreen}{rgb}{0.56, 0.74, 0.56}
\definecolor{darkspringgreen}{rgb}{0.09, 0.45, 0.27}
\definecolor{dollarbill}{rgb}{0.52, 0.73, 0.4}

\newcommand{\delete}[1]{{\color{red}\sout{}}}

\newcommand{\versionnum}[0]{9.0\today~@ 17:40 CET} 
\newcommand{\microsubmissionnumber}{137}

\fancypagestyle{firstpage}{
  
  \fancyhead[C]{\normalsize{HPCA 2019 Submission
  \textbf{\#\microsubmissionnumber} -- Confidential Draft -- Do NOT Distribute!! \\
  \textcolor{MidnightBlue}{\emph{Version \versionnum}}
  }}
  \pagenumbering{arabic}
}

\begin{document}
\bstctlcite{IEEEexample:BSTcontrol} 


\title{FastRemap: A Tool for Quickly Remapping Reads \\between Genome Assemblies}


%


\author{
{Jeremie S. Kim\affilETH}\qquad%
{Can Firtina\affilETH}\qquad%
{Meryem Banu Cavlak\affilETH}\qquad
{Damla Senol Cali\affilBionano}\qquad \\
{Can Alkan\affilBilkent}\qquad 
{Onur Mutlu\affilETH$^{,}$\affilBilkent}\qquad\vspace{-3mm}\\\\%
{\vspace{-3mm}\emph{\affilETH ETH Z{\"u}rich \qquad
\qquad \affilBionano Bionano Genomics \qquad \affilBilkent Bilkent University}}%
}


%


\maketitle
\thispagestyle{plain} 
\pagestyle{plain}

\setstretch{1.01} 
\renewcommand{\footnotelayout}{\setstretch{0.9}}

\camerareadytrue


\fancyhead{}
\ifcameraready
 \thispagestyle{plain}
 \pagestyle{plain}
\else
 \fancyhead[C]{\textcolor{MidnightBlue}{\emph{Version \versionnum~---~}}}
 \fancypagestyle{firststyle}
 {
   \fancyhead[C]{\textcolor{MidnightBlue}{\emph{Version \versionnum~---~}}}
   \fancyfoot[C]{\thepage}
 }
 \thispagestyle{firststyle}
 \pagestyle{firststyle}
\fi




%


\input{abstract}

\input{1_introduction}

\input{2_features_methods}

\input{3_results}

\SetTracking
 [ no ligatures = {f},
 outer kerning = {*,*} ]
 { encoding = * }
 { -40 } 

{

  \footnotesize
  \renewcommand{\baselinestretch}{0.5}
  \let\OLDthebibliography\thebibliography
  \renewcommand\thebibliography[1]{
    \OLDthebibliography{#1}
    \setlength{\parskip}{0pt}
    \setlength{\itemsep}{0pt}
  }
  \bibliographystyle{IEEEtranS}
  \bibliography{ref}
}


\end{document}

%% file: abstract.tex
\begin{abstract}

A genome read data set can be quickly and efficiently remapped from one reference to another similar reference (e.g., between two reference versions or two similar species) using a variety of tools, e.g., the commonly-used CrossMap tool. With the explosion of available genomic data sets and references, high-performance remapping tools will be even more important for keeping up with the computational demands of genome assembly and analysis.

We provide FastRemap, a fast and efficient tool for remapping reads between genome assemblies. FastRemap provides up to a 7.82$\times$ speedup (6.47$\times$, on average) and uses as low as 61.7\% (80.7\%, on average) of the peak memory consumption compared to the state-of-the-art remapping tool, CrossMap.

FastRemap is written in C++. Source code and user manual are freely available at: github.com/CMU-SAFARI/FastRemap. Docker image available at: https://hub.docker.com/r/alkanlab/fast. Also available in Bioconda at: https://anaconda.org/bioconda/fastremap-bio.

%
%


\end{abstract}

%% file: 1_introduction.tex
\section{Introduction} 
\label{sec:intro}

Genome read data sets that have already been mapped to one reference must often
be \emph{remapped} to another reference. This is especially important for cases
where 1)~a more accurate reference genome is released or 2)~the read data set
was previously mapped to a reference genome of the wrong \emph{species}.
Several \emph{remapping tools} (pypi.org/project/segment-liftover~\cite{gao2018segment_liftover}; 
genome.ucsc.edu/cgi-bin/hgLiftOver~\cite{kuhn2013ucsc}; crossmap.sourceforge.net~\cite{zhao2013crossmap};
ncbi.nlm.nih.gov/genome/tools/remap; pypi.org/project/pyliftover)
exist for this purpose and provide a means for remapping a read data set from
one (\emph{source}) reference to another similar (\emph{target}) reference
(e.g., between two reference versions or two similar species).
CrossMap~\cite{zhao2013crossmap} is among the most commonly used
since it supports remapping commonly-used BAM/SAM files as well as many other
data file formats used in genome analysis (e.g., BED, VCF). CrossMap provides
significant speedup compared to fully mapping a read data set to the target
reference using a conventional read mapping (\emph{not} remapping) tool (e.g.,
BWA-MEM~\cite{li2013aligning}), since CrossMap exploits the knowledge of shared
(i.e., identical) genomic regions and their positions within each of the
source/target pair of references in order to quickly update a read's mapping
position.  If a read maps within a shared region of the source reference, the
read will map within the same shared region (with the same offset into the
region) of the target reference, and thus a remapping tool can infer the
read's mapping position in the new reference.

Despite its wide use in the area~\cite{luu2020benchmark}, CrossMap has several
important shortcomings. First, it is not optimized for performance, leaving
much to be desired in terms of remapping speed. We believe speed of remapping
tools is critical due to continued explosion of available genomic data,
frequent improvements to existing references, and the prohibitive computational
power that would be required to completely rely on read mapping tools to
repeatedly map all read data sets to the most recent reference genome versions.
Second, we find that the immediate BAM/SAM output of CrossMap is incompatible
with downstream analysis tools (e.g., GATK HaplotypeCaller~\cite{McKenna2010},
Strelka2~\cite{Kim2018strelka2}, Platypus~\cite{rimmer2013platypus}). Enabling
compatibility is critical to accurately understand the final mapping
results. 

In order to address the key limitations of CrossMap, in this work, we provide
FastRemap, a new C++ implementation of the BAM/SAM remapping component of
CrossMap with key modifications that enable downstream analysis on its
immediate outputs. In our evaluation of FastRemap and CrossMap on three
different sizes of reference genomes (i.e., human, Caenorhabditis elegans, and yeast), we
find that FastRemap provides up to 7.82$\times$ speedup (6.47$\times$, on
average) and uses as low as 61.7\% (80.7\%, on average) of the peak memory
consumption compared to the state-of-the-art remapping tool, CrossMap. 

FastRemap is open source and all scripts needed to replicate the results in
this paper can be found at https://github.com/CMU-SAFARI/FastRemap. FastRemap
is also available at conda in the bioconda
channel as fastremap-bio~\cite{gruning2018bioconda}.

%% file: 2_features_methods.tex
\section{Features and Methods}\label{sec:methods}

To remap reads from one (source) reference to another (target) reference,
FastRemap relies on a chain file (specific to the pair of references), which
indicates regions that are shared between the two references. The chain files
that we use are available on the UCSC genome browser
(hgdownload.soe.ucsc.edu/downloads.html).  We model our codebase off of
CrossMap, such that it is easily extensible to other data file formats. We
currently support remapping reads in the most commonly used formats (i.e.,
SAM/BAM, BED).  However, we plan to extend support for additional file formats
(e.g., VCF, GTF/GFF, BigWig, MAF). 

\noindent\textbf{Libraries.}
We use the Seqan2 library (Knut et al., 2017) for manipulating genomic data and
zlib (github.com/madler/zlib) for enabling the input and output of compressed
files (i.e., BAM format). 

\noindent\textbf{Upgrades to the CrossMap Implementation.} In addition to
providing a faster C++ implementation, we make two key changes to the CrossMap
code logic to enable accurate and immediate downstream analysis on FastRemap's
output. First, CrossMap removes the supplementary alignment flag when remapping
BAM/SAM files. These incorrect flag values cause Picard's MarkDuplicates tool
(gatk.broadinstitute.org/hc/en-us/articles/360037052812-MarkDuplicates-Picard-)
to fail, which prevents further downstream analysis beyond the MarkDuplicates
step (e.g., GATK HaplotypeCaller for variant calling). FastRemap correctly sets
the supplementary alignment flag when remapping BAM/SAM files, enabling
downstream analysis tools to correctly process the results. Second, FastRemap
outputs reads that have failed to remap into a separate BED file, which can be
referenced for further analysis. This allows the user to immediately identify
and process these reads from the BED file while minimizing the disk space
requirements for storing these reads.

%% file: 3_results.tex
\section{Results}

\begin{figure}[!t]
    \centering
    \includegraphics[width=\columnwidth]{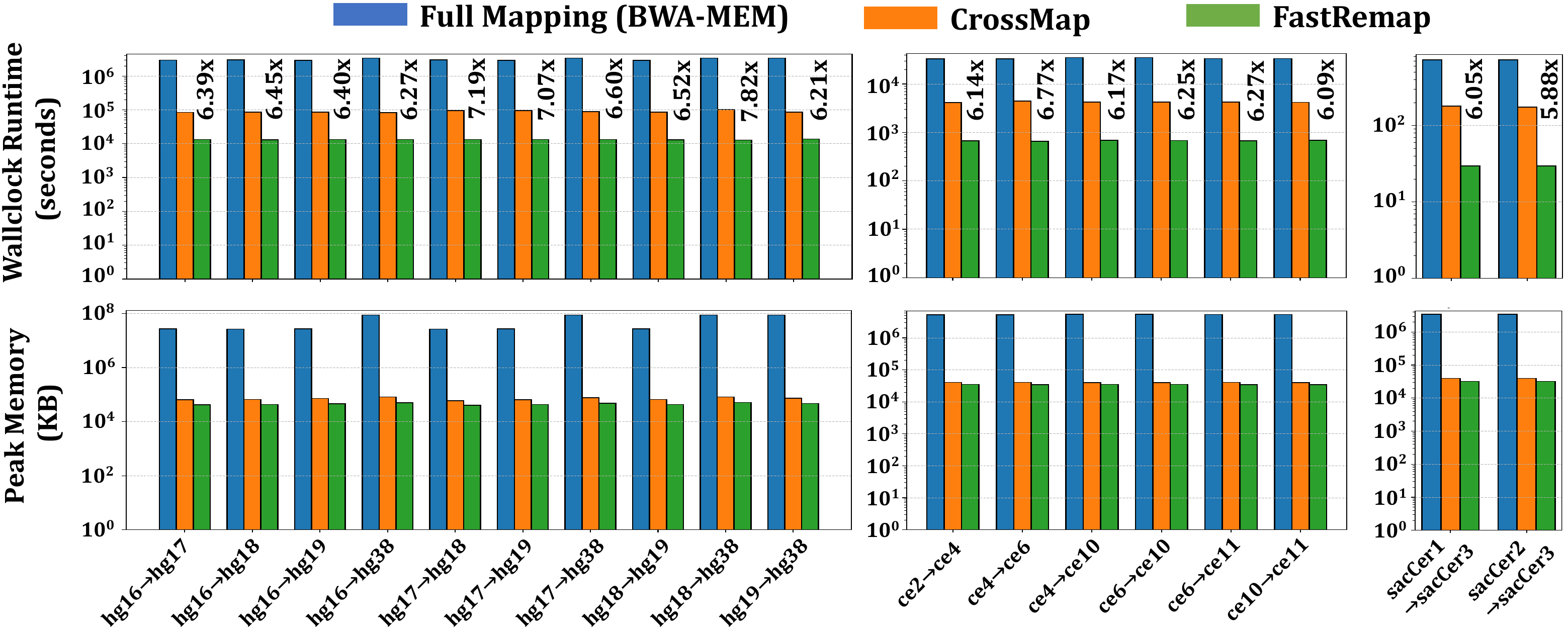}
    \caption{Performance (wall clock runtime) and peak memory usage evaluation of Full Mapping with BWA-MEM, FastRemap, and CrossMap. We show speedup values between FastRemap and CrossMap above each set of bars in the chart.} 
    \label{fig:evaluation}
\end{figure}

Before discussing our evaluation results of peak memory consumption and performance, we describe our methodology for evaluation. 

\subsection{Evaluation Methodology} 

\noindent\textbf{Remapping Tools.} We compare FastRemap to the state-of-the-art
remapping tool CrossMap and also full mapping with BWA-MEM. 

\noindent\textbf{Evaluated Reference Genomes.} We evaluate FastRemap with several versions of reference genomes of varying size across 3 species (i.e., human, C. elegans, yeast). 

\noindent\textbf{Evaluated Read Data Sets.} We use publicly available DNA-seq read sets for each reference under examination, which are summarized in Table~\ref{tab:read_data_sets}.

\begin{table}[!t]
\caption{Read datasets we use in our evaluations.}\label{tab:read_data_sets} {\begin{tabular}{@{}lll@{}}\toprule Read Data Set & Accession No. & No. Paired-end reads \\\midrule
Human NA12878 & ERR194147  & 791,385,507 (101bps) \\
Human NA12878 & ERR262997  & 643,097,275 (101bps) \\
C. elegans N2 & SRR3536210 & 78,696,056 (101bps)  \\
Yeast S288C   & ERR1938683 & 3,318,467 (150bps)   \\\toprule
\end{tabular}}{} 
\end{table}


\noindent\textbf{Evaluation System.} We run FastRemap on a state-of-the-art server with 64 cores (2 threads
per core, AMD EPYC 7742 @ 2.25GHz), and 1TB of main memory. We collect the end-to-end wallclock runtime (i.e., total execution time from the program's start to finish from the perspective of the user) and memory usage using the $time$ command in Linux with $-vp$ flags. We report the \emph{wallclock runtime} (in seconds) and \emph{peak memory usage} (in megabytes) of our evaluations based on these configurations.

\subsection{Evaluation Results} \label{subsec:evresult}

Figure~\ref{fig:evaluation} shows the evaluation results of FastRemap and
CrossMap, as compared to full mapping using BWA-mem for wallclock runtime (Fig.
1, top graph) and peak memory usage (Fig. 1, bottom graph) across various pairs
of reference versions for human (\emph{hg}), C. elegans (\emph{ce}), and yeast
(\emph{sacCer}). The reference pairs are denoted as
\emph{source}$\rightarrow$\emph{target} on the shared x-axis. 

We make three key observations. 
First, FastRemap provides significant wallclock speedup (i.e., between 5.88$\times$ and 7.82$\times$) compared to CrossMap. This speedup mainly comes from FastRemap's utilization of 4 threads in the Seqan code segments, whereas CrossMap's code is purely single-threaded. Second, FastRemap requires a smaller peak memory footprint (between 61.7\% and 88.0\% of that requuired by CrossMap) compared to
CrossMap's, which is likely due to the smaller data structures that are used by C++ compared to Python. FastRemap's lower memory usage provides higher scalability than CrossMap when running many instances in parallel. Third, compared to fully mapping the read set using BWA-MEM, FastRemap is significantly faster (150.3$\times$ on average) and much more memory efficient (requiring only 0.4\% of the memory footprint of BWA-MEM on average).

We conclude that FastRemap provides significant performance, memory, and functional benefits over the state-of-the-art CrossMap tool when remapping BAM/SAM files. We hope that FastRemap enables academia and industry to develop powerful analysis pipelines for quickly studying the effects of mapping a read data set to a different
reference genome (e.g., AirLift~\cite{kim2019airlift}).